\documentclass[superscriptaddress,prl,twocolumn,preprintnumbers,amsmath,amssymb]{revtex4-1}
\usepackage{epsfig}
\usepackage{amsmath}
\usepackage{array}
\usepackage{multirow}
\usepackage{psfrag}
\usepackage[font=footnotesize,format=plain,labelsep=period,justification=raggedright]{caption}
\usepackage{subcaption}
\usepackage{graphicx}
\usepackage{color}
\usepackage{comment}

\begin{document}

	\title{Phase Separation in Ternary Fluids}
	\author{Alvin C. M. Shek}
	\affiliation{Department of Physics, Durham University, Durham, DH1 3LE, UK}
	\author{Halim Kusumaatmaja}
	\email{Email: halim.kusumaatmaja@durham.ac.uk}
	\affiliation{Department of Physics, Durham University, Durham, DH1 3LE, UK}
	\date{\today}
	
	\begin{abstract}
		We computationally study the spontaneous phase separation of ternary fluid mixtures using the lattice Boltzmann method both when all the surface tensions are equal and when they have different values. To rationalise the phase diagram of possible phase separation mechanisms, previous theoretical works typically rely on analysing the sign of the eigenvalues resulting from a simple linear stability analysis, but we find this does not explain the observed simulation results. Here, we classify the possible separation pathways into four basic mechanisms, and develop a phenomenological model that captures the composition regimes where each mechanism is prevalent. We further highlight that the dominant mechanism in ternary phase separation involves enrichment and instability of the minor component at the fluid-fluid interface, which is absent in the case of binary fluid mixtures.
	\end{abstract}
	\maketitle
	\section{Introduction}
	Spontaneous phase separation of fluid mixtures is important for a wide range of processes. For example, in nature, it is known to give rise to structural colours in living creatures, such as observed in birds and butterflies \cite{narasimhan2018multifunctional,burg2018self}. More recently, it has also been shown to play fundamental roles in volcanic eruptions \cite{allabar2018message,sahagian2020explosive} and in the organisation of cellular matters, leading to the formation of the so-called biomolecular condensates or membraneless organelles \cite{alberti2017phase,boeynaems2018protein,shin2017liquid,Kusumaatmaja2021}. In industry, better understanding of the phase separation process is necessary, among others, for tuning the formation of fractures in alloys in the field of metallurgy \cite{cahn1961spinodal,cahn1965phase}, for manipulating the structure of polymer blends which in turn affect their mechanical and electrical properties \cite{goldsbrough1972spinodal,si2015liquid,kuang2013defect}, and for controlling the morphology of complex emulsions for applications in drug delivery \cite{de2010polymeric,de2012polymeric,krawczyk2016elasticity} and in the food industry \cite{guzey2006formation,benjamin2012multilayer,burgos2016multilayer}.
	
	Extensive theoretical and experimental studies on phase separation have been carried out in the case where the fluid mixtures separate into two distinct, immiscible fluid components \cite{verhaegh1996fluid,wagner1998,wagner1999phase,patzold1995numerical,tanaka2001interplay,matas2014hydrodynamic,kumar2019localization,das2020surface}. For such binary fluid case, it is now generally well understood when spontaneous phase separation occurs, what the resulting morphologies are, how the separated domains coarsen (both with and without the influence of hydrodynamics), and how the process may be affected by the patterning of the solid boundary. However, the general problem of phase separation is significantly more complex, and there are numerous instances where the fluid mixtures separate into more than two immiscible components \cite{mao2019phase,mao2020designing,Jacobs2013,Jacobs2017}. These scenarios, in contrast, have received less attention and remain poorly understood.    
	
	Our focus in this work is on phase separation of three immiscible fluid components. In the literature, this ternary fluid case has primarily been studied in the context of thin films of polymer blends \cite{wang2000thin,chung2004mobile,reich1981phase} where the evolution of their morphologies for a number of specific polymer compositions have been tabulated experimentally \cite{nauman1994morphology}. There is also growing interest in ternary fluid phase separation for other applications, such as a novel route for the production of complex droplet emulsions, nanoparticles and patchy droplets \cite{chao2018generation,zhang2021phase,park2020photonic,ekanem2017facile}. In addition, from the modelling side, there have been efforts to simulate phase separation pathways that reproduce the experimental observations \cite{nauman1994morphology,chen1994computer,huang1995phase,Huang1996,Smith2000}, including scaling analysis on the domain coarsening  \cite{tanaka1998spontaneous,rogers1989numerical,tateno2021power,Singh2015}.  Yet, despite these advances, there is still limited understanding in one of the most fundamental aspects of ternary fluid phase separation: how to predict and characterise the different possible distinct morphologies and phase separation pathways as function of the fluid composition. To provide insights, surprisingly, works to date have primarily relied on a simple linear stability analysis to demarcate the ternary phase diagram into regions with zero, one and two positive eigenvalues \cite{nauman1994morphology,chen1994computer,huang1995phase,sugathan2020phase}. As we will demonstrate here, considering only the sign of the eigenvalues do not allow qualitative, let alone quantitative, predictions for the separation pathways. 
	
	Our contribution here is three-fold. First, in agreement with previous literature, we observe numerous possible morphologies and separation pathways. Here, we group them into 4 distinct categories, and rationalise these groupings by extending the prevailing linear stability analysis and harnessing information provided by the resulting eigenvalues and eigenvectors. 
	Second, we highlight that the dominant mechanism across the composition phase space in ternary phase separation is where the minor fluid component is enriched and undergoes an instability at the interfaces between the two more major components. Third, we show that the theoretical framework can be applied when the interfaces all have the same surface tension, as well as when they all take different values.
	
	
	\section{Lattice Boltzmann Simulation Method}
	To simulate the ternary phase separation dynamics, we employ a free energy lattice Boltzmann method. The detailed lattice Boltzmann implementation is similar to the one described by Semprebon {\it et al.} \cite{semprebon2016ternary}.
	Here we only summarise the most salient aspects of the continuum model we employ.
	
	The free energy model we use is based on the work of Boyer and Lapuerta \cite{boyer2006study},
	\begin{align}
	F & = \int \sum_{m=1}^{3}\frac{12}{\varepsilon}\frac{\Sigma_m}{2}\left( C_m^2(1-C_m)^2 + \frac{\varepsilon^2}{16} (\nabla C_m)^2\right) dV.
	\end{align}
	This model implicitly assumes we are below the critical temperature where three fluid components can coexist. Here, the concentration or volume fraction for each component is given by $C_m$. The coefficients $\Sigma$'s are related to the interfacial tensions, such that the interfacial tension between fluids $m$ and $n$ is given by $\gamma_{mn} = (\Sigma_m + \Sigma_n)/2$. The coefficient $\varepsilon$ describes the characteristic interface width, which is taken to be the same for all fluid interfaces. 
	
	The fluid equations of motion that we solve correspond to the Cahn-Hilliard, continuity and Navier-Stokes equations. The Cahn-Hilliard equations capture the evolution of the fluid interfaces, and for each component $C_m$, it is given by 
	\begin{align}
	& \partial_t C_m + \partial_{\alpha}(C_m u_\alpha)  = \partial_{\alpha} \left(M_m \partial_{\alpha} \mu_m\right). \label{Eqn:dC}
	\end{align}
	Here, $\mathbf{u}$ is the fluid velocity, $M_m = M_o/\Sigma_m$ is the mobility parameter for each fluid component $m$ ($M_o$ is a constant), and $\mu_m$ is the chemical potential. Following the work by Boyer and Lapuerta \cite{boyer2006study},
	\begin{align}
	\mu_m  = \frac{\delta F}{\delta C_m} + \beta = -\frac{3}{4} \varepsilon \kappa_m\nabla^2 C_m + \frac{12}{\varepsilon} \Sigma_m H(C_m) + \beta,
	\end{align}
	where
	\begin{eqnarray}
	& H(C_m) = C_m(1-C_m)(1-2C_m), \\
	& \beta = - \frac{12}{\varepsilon}\delta C_1 C_2 C_3, \\
	& \delta = \dfrac{6\Sigma_1\Sigma_2\Sigma_3}{\Sigma_1\Sigma_2+\Sigma_1\Sigma_3+\Sigma_2\Sigma_3}.
	\end{eqnarray}
	The $\beta$ term is a Lagrange multiplier term to ensure that the constraint $C_1+C_2+C_3 = 1$ is maintained. The introduction of this term leads to two desirable properties. First, the model becomes reducible, meaning that when one component is absent from the system, then the resulting equations of motion become independent of the removed component. Second, it inhibits unphysical growth of a third fluid component at the interface between two fluids. This can occur when the model is not reducible.
	
	The Cahn-Hilliard equations are coupled to the continuity and Navier-Stokes equations that describe the hydrodynamics of the fluids:
	\begin{eqnarray}
	\partial_t \rho + \partial_{\alpha}(\rho u_{\alpha}) &=& 0, \\
	\partial_t (\rho u_{\alpha}) + \partial_{\beta}(\rho u_{\alpha} u_{\beta}) &=& -\partial_{\beta}P_{\alpha\beta} + \nonumber \\
	&& \partial_{\beta}\eta(\partial_{\beta}u_{\alpha} + \partial_{\alpha}u_{\beta}).
	\end{eqnarray}
	In this work, for simplicity, we have assumed all components have the same density, such that the total density $\rho = C_1+C_2+C_3 = 1$ is constant. The pressure tensor is related to the chemical potential via
	\begin{align}
	\partial_{\beta}P_{\alpha \beta} = \partial_{\alpha} p + \sum_{m=1}^{3} \left[C_{m}\partial_{\alpha}\mu_m\right].
	\end{align}
	As is standard in the lattice Boltzmann method \cite{kruger2017lattice,semprebon2016ternary}, the ideal gas pressure contribution is $p=\rho c_s^2$, with the speed of sound $c_s = 1/\sqrt{3}$.
	
	In all situations studied in this work, we initialise each simulation by introducing small random concentration perturbations (white noise) on top of a homogeneous mixture at a given composition, typically with an amplitude of $10^{-4}$ of the fluid density. In this range, we verify the simulation results do not sensitively depend on the choice of noise amplitude. We have also systematically varied the size of our simulation box, and find a domain size of 240 x 240 is suitable to robustly identify the different phase separation pathways for constructing the phase diagrams.
	
	For all the simulations explicitly presented here, the following parameters are set to be $M_0 = 0.005$, $\varepsilon = 4.0$, and $\eta = 0.167$. For the surface tension, we use $\Sigma_1 = \Sigma_2 = \Sigma_3 = 0.0133$ in the equal surface tension case. When we vary the surface tension, we set $\Sigma_1 = 0.0133$, $\Sigma_2 = 9 \Sigma_1$ and $\Sigma_3 = 4 \Sigma_1$.
	
	\section{Equal Surface Tension Case}
	
	We first focus on the simplest scenario where the surface tensions between any two pair of liquids are equal. We will consider the case  where the surface tensions are different in the next section. 
	
	\subsection{Morphologies and Pathways}
	Let us start by considering the structures which can occur as the ternary fluid mixtures undergo phase separation. We find there are four distinct types of separation pathways, which typically lead to four morphology classes. 
	
	\begin{figure} [h]
		\centering
		\includegraphics[width=\linewidth]{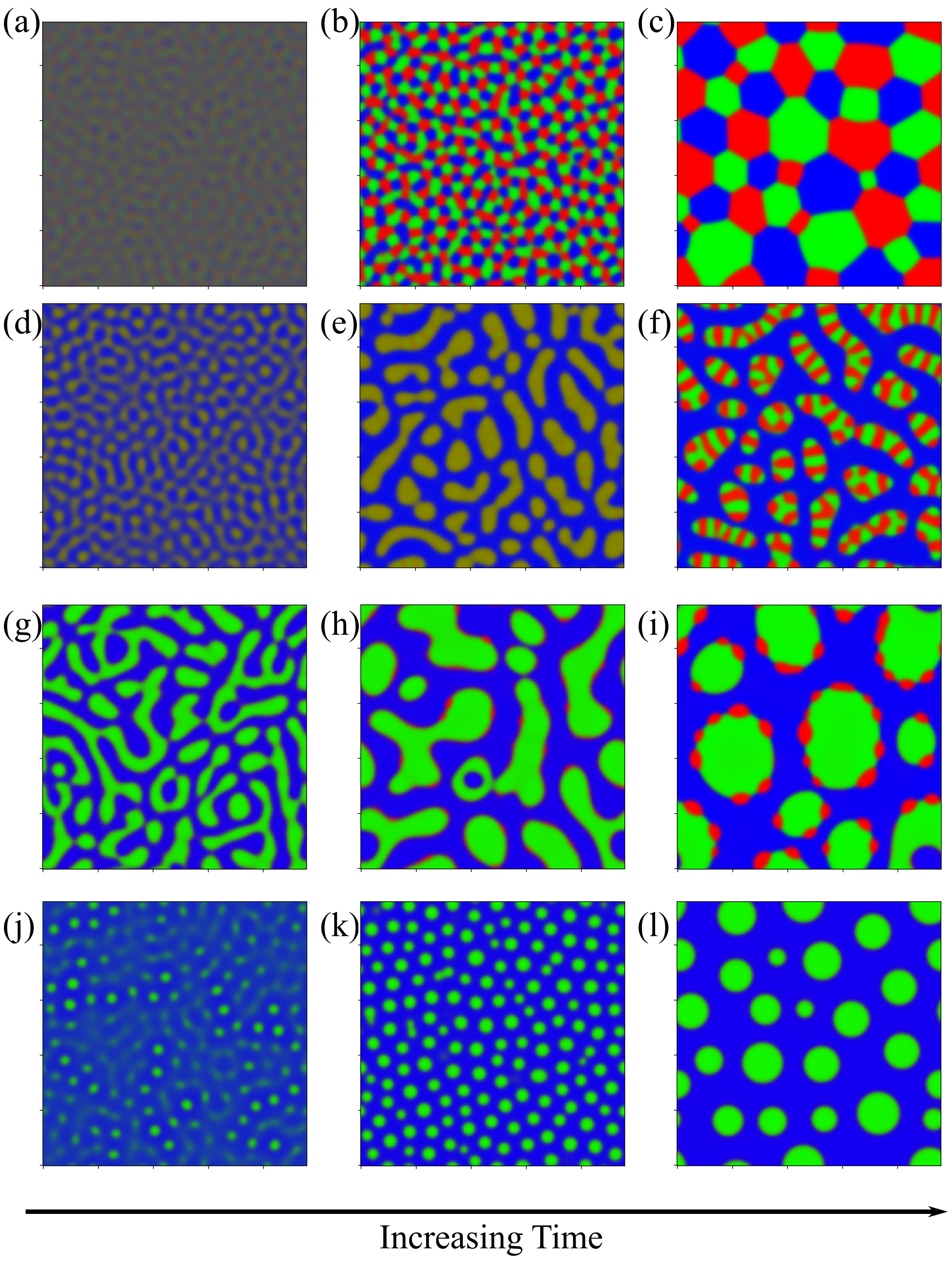}
		\caption{\label{Separation_Pathways} Four types of spontaneous phase separation pathways for ternary fluid mixtures: (a-c) direct ternary separation, (d-f) primary and secondary bulk phase separation, (g-i) enrichment and instability at interface, and (j-l) binary-like phase separation.}
	\end{figure}
	
	Type I of ternary phase separation is when all three fluid components begin to simultaneously separate. This is exemplified in Fig. \ref{Separation_Pathways}(a-c), leading to what we term a lattice morphology. Here, separate domains corresponding to the three fluids are interspersed among each other, and domain coarsening occurs due to rearrangement and coalescence of alike domains. 
	
	For types II and III of ternary phase separation, the pathways consist of two stages, in contrast to only one stage for type I above. Fig. \ref{Separation_Pathways}(d-f) illustrate type II where primary and secondary bulk separation (spinodal decomposition) occur consecutively. This most commonly leads to what we term the worm morphology, where there is a chain of alternating fluid domains. During the primary spinodal (panel e), two components (e.g. without any loss of generality, $C_1$ and $C_2$) remain mixed and together they separate out from the third (e.g. $C_3$) component. During the secondary bulk phase separation (panel f), the two initially mixed components undergo another spinodal decomposition.
	
	An example of type III of ternary phase separation is shown in Fig. \ref{Separation_Pathways}(g-i). As we will show in the phase diagrams in the following sections, this is the most occuring pathway. Unlike for the binary phase separation where spinodal decomposition is the sole driving mechanism for spontaneous phase separation, in ternary fluid case, we have an alternative mechanism. Here, the primary spinodal is followed by the enrichment of the minority component at the interface (panel h). When there is sufficient third component at the interface, this component eventually becomes unstable and form small droplets at the fluid-fluid interface (panel i). This type of ternary phase separation typically gives rise to patchy droplet morphology.
	
	Finally, for type IV, as in Fig. \ref{Separation_Pathways}(j-l), when there is insufficient minor component at the interface, the phase separation pathway is akin to the binary fluid case. Here, the smallest minority component is never fully phase separates but it is typically concentrated at the interface between the two more major components.
	
	Our simulation results are fully consistent with previous works describing possible phase separation pathways (sometimes also termed as decomposition patterns) in ternary fluid mixtures \cite{nauman1994morphology,chen1994computer,huang1995phase,Huang1996,Smith2000}. Many of the extensive morphologies previously tabulated are a result of stochastic collision and coalescence of different fluid structures at the later stages of the coarsening dynamics. For example, Nauman and He recorded 27 possible morphologies \cite{nauman1994morphology}. In contrast, here we focus on the instability mechanisms and simplify the classification into 4 basic types of phase separation mechanisms. We find these same 4 instability types when we vary the fluid viscosity or when we simulate the purely diffusive regime (by turning off coupling to hydrodynamics).
	
	
	\subsection{Phase Diagram}
	
	Building on the four fundamental phase separation mechanisms identified in the previous sub-section, a natural question arises: can we predict which pathway and morphology will form given the mixture composition? In the literature to date, the phase diagram for ternary phase separation is commonly interpreted by performing a linear instability analysis and looking into the eigenvalues. This is a standard tool to understand phase separation via spinodal decomposition \cite{cahn1961spinodal,cahn1965phase,huang1995phase,siggia1979late}. 
	
\begin{figure*}
		\centering
		\includegraphics[width=1.0\linewidth]{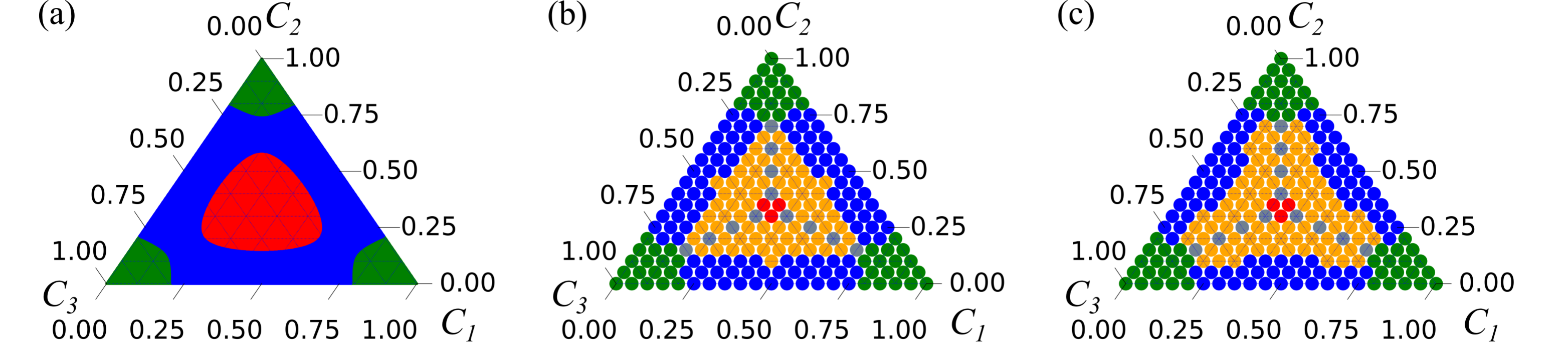}
		\caption{\label{PhDiagram} Phase diagrams obtained (a) from considering the signs of the eigenvalues from linear instability analysis, (b) from lattice Boltzmann simulation results, and (c) from the phenomenological model proposed in this work. In panel (a), the red, blue and green regions have two, one, and zero positive eigenvalues. In panels (b) and (c), red denotes the direct ternary phase separation mechanism, gray the primary and secondary bulk phase separation mechanism, orange the enrichment and instability at the interface mechanism, blue the binary-like mechanism, and green no phase separation.}
	\end{figure*}

	Suppose the system is initialised as a homogeneous mixture with small perturbations in composition, such that, in 1-D,
	\begin{align}
	C_m(x,t = 0) = A_m + \sum_{k}^{\infty}\cos(kx) \alpha_{mk}(t = 0),
	\end{align}
	where the $A_m$'s are the initial concentrations, and the $\alpha_{mk}(t)$'s describe the amplitudes of the perturbations. If we substitute the above equation into the Cahn-Hilliard equation, Eq. \ref{Eqn:dC}, with zero initial velocity, to linear order we obtain (the prime denotes a time derivative)
	\begin{align}
	\begin{pmatrix}
	\alpha_{1k}^{\prime}(t) \\
	\alpha_{2k}^{\prime}(t) 
	\end{pmatrix} 
	= 
	\begin{pmatrix}
	D_1 + F_{11} & F_{12} \\
	F_{21} & D_2 +F_{22}
	\end{pmatrix} 
	\begin{pmatrix}
	\alpha_{1k}(t) \\
	\alpha_{2k}(t) 
	\end{pmatrix} 
	\end{align}		
	with 
	\begin{align}
	D_i & = \frac{-3}{4} \varepsilon k^4 - \frac{12}{\varepsilon} (6A_i^2 - 6A_i + 1) k^2, \\
	F_{i1} & = \frac{12\delta}{\varepsilon \Sigma_i}(A_2(1-2A_1-A_2)) k^2, \label{eq:Fi1}\\
	F_{i2} & = \frac{12\delta}{\varepsilon \Sigma_i}(A_1(1-2A_2-A_1)) k^2. \label{eq:Fi2}
	\end{align}
	Without any loss of generality, we have explicitly considered variations in $C_1$ and $C_2$. The third concentration can be obtained by using the constraint $C_3 = 1 - C_1 - C_2$. From the matrix equation, the eigenvalues can then be determined:
	\begin{align} \label{eigenvalues}
	\lambda^{\pm} & = \frac{-b \pm \sqrt{b^2 - 4c}}{2}, \\
	b & = -(D_1 + F_{11} + D_2 + F_{22}), \\
	c & = (D_1 + F_{11})(D_2 + F_{22}) - F_{12} F_{21}.
	\end{align}
	The eigenvalues can be used to differentiate between three regions in the phase diagram: (i) where both eigenvalues are negative, (ii) where one is positive and one negative, and (iii) where both eigenvalues are positive. These regions are shown in Fig. \ref{PhDiagram} (a) in green, blue and red respectively. In our phase diagrams the corners of the triangle correspond to when one component is maximal, i.e. $C_1 = 1.0$ for the bottom right corner, $C_2 = 1.0$ for the top corner, and $C_3 = 1.0$ for the bottom left corner.
	
	Numerically, we can also construct a phase diagram by varying the initial fluid composition and observing the resulting fluid structures in the simulations. This phase diagram is shown in Fig. \ref{PhDiagram} (b). Here, we have used red to represent direct ternary phase separation, gray for two-step primary and secondary spinodal decomposition, orange for enrichment and instability at interface, and blue for binary-like phase separation. When comparing the the two phase diagrams in Fig. \ref{PhDiagram} (a) and (b), we can immediately conclude that knowing only the eigenvalues is far from adequate for predicting ternary fluid phase separation. Extending the linear stability analysis to higher dimension (2D) also will not capture the discrepancy we have highlighted. Fundamentally, both the primary and secondary bulk phase separation and the enrichment and instability at the interface mechanisms are two step processes that cannot be captured from a linear stability analysis simply using information on the initial mixture concentrations. Hence, our next aim is to develop a simple phenomenological model that better describe the numerical phase diagram shown in Fig. \ref{PhDiagram}(b). 
	
	From the numerical results, while the red region in Fig. \ref{PhDiagram}(a) with two positive eigenvalue covers a significant area, we find the direct ternary phase separation scenario is very limited. Across all surface tension values we have studied, direct ternary phase separation is limited to compositions where all the concentrations satisfy $\frac{1}{3} \pm 5\%$. This is the region marked as red in Fig. \ref{PhDiagram}(c). Geometrically, the lattice morphology is favoured for such composition due to packing constraint as there simply is not enough space for the other types of morphologies to form.  
	
	Next, we will study the two-step primary and secondary spinodal decomposition scenario. To do this, it is useful to consider the full linearised solution for the evolution of the perturbation amplitude, given by
	\begin{align}
	\underline{\alpha}(t) = B^+ \exp^{\lambda^+ t}\underline{e}^+ + B^- \exp^{\lambda^- t}\underline{e}^-,
	\end{align}
	where $B^\pm$ are fixed by the initial conditions, and $\underline{e}^{\pm}$ are the two eigenvectors corresponding to eigenvalues $\lambda^{\pm}$, as given in Eq. \eqref{eigenvalues}. For equal surface tension considered in this section, when $A_1 = A_2 = A$, we find $D_1 = D_2 = D$ and $F_{i1} = F_{i2} = F$, and the full linearised solution simplifies to 
	\begin{align}
	\underline{\alpha}(t) = B^+ \exp^{\lambda^+ t}\begin{pmatrix}
	1 \\ 1
	\end{pmatrix} + B^- \exp^{\lambda^- t}\begin{pmatrix}
	1 \\ -1
	\end{pmatrix} \label{eigenvector}
	\end{align}
	with eigenvalues
	\begin{align}
	\lambda^{\pm} = D+2F, \hspace{0.5 cm} D.
	\end{align}
	The $\lambda^-$ mode is the standard binary mode of separation where $D>0$ leads to unstable situation. Here, the perturbations in components $C_1$ and $C_2$ grow with opposite signs. In contrast, for the $\lambda^+$ mode, the two components grow together at the same rate. Looking at the equations for $F$ in Eqs. \ref{eq:Fi1} and \ref{eq:Fi2}, its sign depends on a term proportional to $(1-3A)$. Hence, if $A<\frac{1}{3}$, $F > 0$, and $\lambda^{+}>\lambda^-$. This condition corresponds to the gray points in Fig. \ref{PhDiagram}(c). In fact, this condition should be a continuous line, but it appears discrete due to how we sample the phase diagram for consistency with the full numerical results in Fig. \ref{PhDiagram}(b).
	
	Dominant $\lambda^{+}$ mode explains the primary spinodal observed in Fig. \ref{Separation_Pathways}(d-f), where two components grow together for a significant period and separate from the third fluid component. If the initial condition satisfies $A_1 + A_2< A_3$, as is the case in Fig. \ref{Separation_Pathways}(d-f), the third component forms the background, while we observe droplets composed of $C_1$ and $C_2$ components. In contrast, if $A_1 + A_2 > A_3$, we find the opposite with $C_3$ droplets and a continuous phase of $C_1$ and $C_2$ mixture. However, these mixed domains cannot grow together indefinitely. At some point, they will enter the spinodal region for the binary mixture and the two components will undergo the secondary bulk phase diagram. As we step off from the $A_1 = A_2 <\frac{1}{3}$ line, we can derive that the eigenvectors do not support any two components to co-grow at the same rate. This limits the region in the ternary phase diagram that phase separates via the two step spinodal pathway.
	
	For the rest of the phase diagram where there is at least one positive eigenvalue, considering the full linear solution leads to the conclusion that ternary phase separation is dominated by the two more major components, in agreement with the numerical results. Further, from energetic arguments, it is favourable for the minor component to be enriched at the interface \cite{Huang1996,Smith2000}, instead of uniformly diffused in the bulk of the major components. The remaining task in this case is to understand if and how the instability at interface takes place. 
	
\begin{figure}
		\centering
		\includegraphics[width=\linewidth]{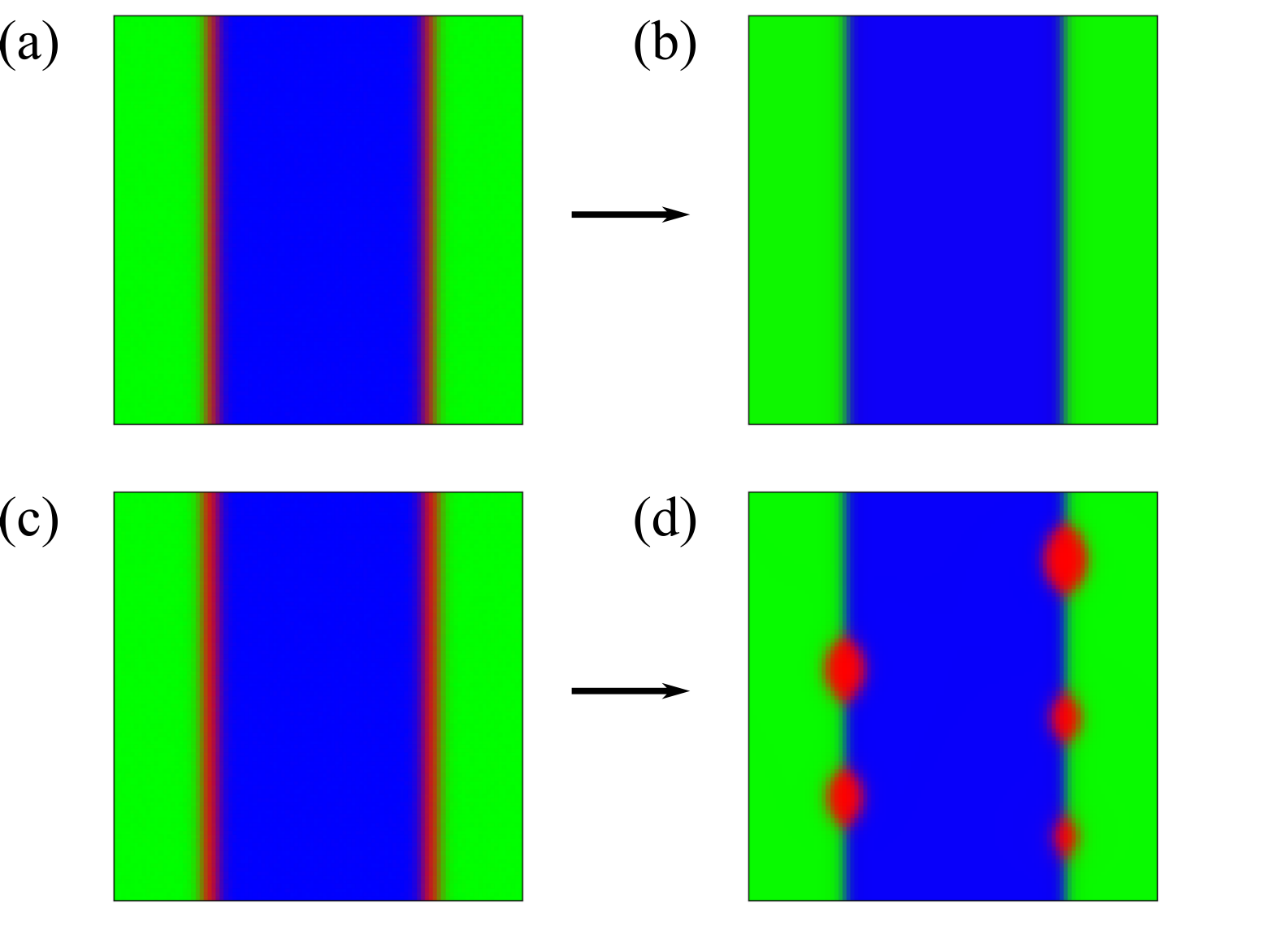}
		\caption{Evolution of the minority component initialised at the interface for (a-b) $a = 3\varepsilon/8$ and (c-d) $a=\varepsilon/2$.}
		\label{TanhStrips}
	\end{figure}

	To study this we simulate fluid strips in 2-D initialised with the following concentration profile, as shown in Fig. \ref{TanhStrips}:
	\begin{align}
	C_1(r) & = 1 - C_2(r) - C_3(r), \\
	C_2(r) & = \frac{1}{2}\left(1+\tanh\left(\frac{2(r-R+a)}{\varepsilon}\right)\right), \\
	C_3(r) & = \frac{1}{2}\left(1-\tanh\left(\frac{2(r-R-a)}{\varepsilon}\right)\right), 
	\end{align}	
	where $r$ is the distance from the centre of a fluid strip, $R$ is half the thickness of the strip, and $a$ is the shift from an equilibrium hyperbolic tangent profile for a strip of $C_2$ surrounded by $C_3$. With increasing $a>0$, we effectively introduce additional minority fluid component $C_1$ at the interface between $C_2$ and $C_3$. Numerically we find the minority component at the interface is stable for small $a$, Fig. \ref{TanhStrips}(a-b), but it becomes unstable for large $a$, Fig. \ref{TanhStrips}(a-b). The transition occurs as $a\sim \epsilon/2$, irrespective of the thickness of the strip. Hence, this instability is primarily determined by the amount of the minority component at the interface, relative to the interface length between the two major components. In addition, since this mode of phase separation requires an instability to form droplets, it is only possible when the spatial dimension is greater than or equal to two. It cannot occur in 1-D. 
	
	With this observation we can make a simple phenomenological model for when patchy droplets will occur.  We first determine when spinodal decomposition will occur, i.e. when at least one of the eigenvalues is positive. Then we consider the cases where type I and II do not occur.  In these cases there are two possibilities left for phase separation, either following type III or IV pathway. 
	
	To predict the boundary between types III and IV, we make the following approximations. We assume that the minor component is evenly distributed at the interfaces with width $w$. Without any loss of generality, here we make the assumption that $C_1 < C_2 < C_3$, such that the minor component $C_1$ is enriched between the $C_2$ droplets and the $C_3$ surrounding. Following the results shown in Fig. \ref{TanhStrips}, we expect instability leading to droplets formation to occur above $w \sim 2a \sim \varepsilon$.
	
	If phase separation between the two major components $C_2$ and $C_3$ continues indefinitely, the total interface length will decrease and $w$ will increase monotonically with time. Hence, eventually we can expect the enrichment and instability mechanism to take place. However, in our simulations, we often observe meta-stable states where the phase separated droplets are well separated and they do not coalesce further. This limits the decrease in the interface length and the increase in $w$. If the number of such droplets is $N$ and the characteristic radius of the $C_2$ droplets is $r_d$, then we can write $V_2 = \pi r_d^2$ and $V_1 + V_2 = \pi (r_d + w)^2$, with $V_1$ and $V_2$ corresponding to the total volume of $C_1$ and $C_2$ components in the system. Combining these relations with the condition $w \sim \varepsilon$, we expect the enrichment and instability mechanism to be observed when
	\begin{align}
	w = \sqrt{\frac{A_2}{\pi n}+\frac{A_1}{\pi n}}- \left(\frac{A_2}{\pi n}\right)^{\frac{1}{2}} \sim \varepsilon. 
	\end{align}
	Here $A_1 = V_1/V_T$ and $A_2 = V_2/V_T$ are the average concentrations of the $C_1$ and $C_2$ components in the system, 
	and $n = N/V_T$ is the typical density of isolated droplets, with $V_T$ the total volume of the simulation domain. Fitting $n$ to obtain the best comparison between the phase diagrams in Fig. \ref{PhDiagram}(b-c), we obtain $n \sim 1.7 \text{ x} 10^{-4}$. In practice, the number of isolated droplets observed does vary depending on the initial concentrations. However, they are in the same order as the fitted value of $n$. Importantly, as $n$ is an intensive variable, our argument holds independent of the system size.
	\begin{figure*}
		\centering
		\includegraphics[width=\linewidth]{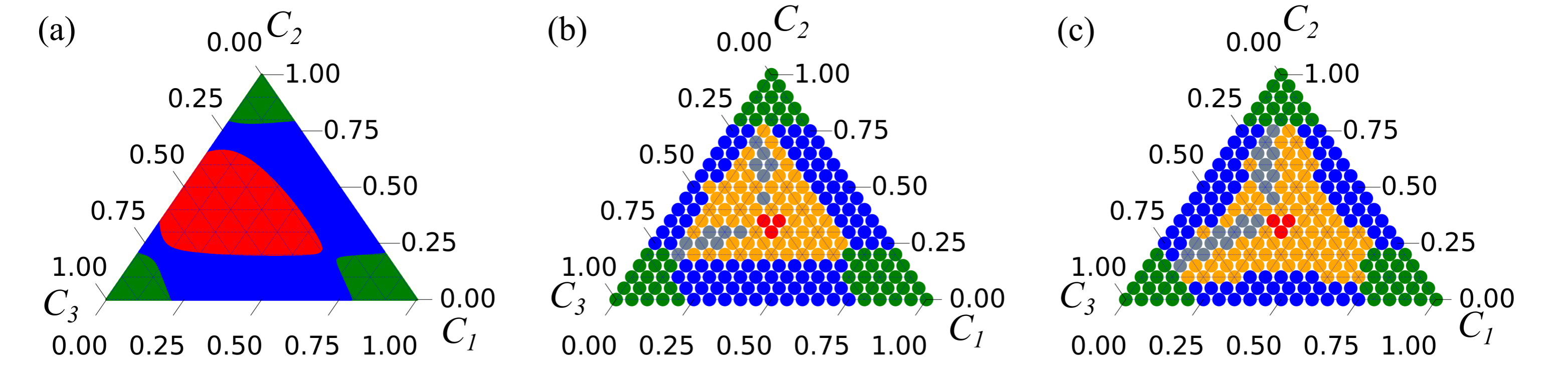}
		\caption{\label{PhaseDiagrams}Phase diagrams for non equal surface tensions where $\Sigma_3 = 9\Sigma_1$, $\Sigma_2 = 4\Sigma_1$. In panel (a), the red, blue and green regions have two, one, and zero positive eigenvalues. Panel (b) is the numerically obtained phase diagram, while panel (c) is our theoretical prediction. In panels (b) and (c), red denotes the direct ternary phase separation mechanism, gray the primary and secondary bulk phase separation mechanism, orange the enrichment and instability at the interface mechanism, blue the binary-like mechanism, and green no phase separation.}
	\end{figure*}
	
	\section{Non-Equal Surface Tensions}
	Having elucidated the case for equal surface tensions, we now generalise our argument for the non equal surface tensions case. As a representative example, we choose a case where the surface tensions are considerably distinct between one another, corresponding to $\Sigma_3 = 4\Sigma_1$, $\Sigma_2 = 9\Sigma_1$ in the free energy model, such that $\gamma_{23}/\gamma_{12} = 1,3$ and $\gamma_{13}/\gamma_{12} = 0.5$. This set of parameters leads to the formation of Neumann angles of $\theta_1 = 63.9^{\circ}$, $\theta_2 = 159.8^{\circ}$, and $\theta_3 = 136.3^{\circ}$ at the three phase contact line. 
	
	
	A major difference for non-equal surface tensions is the occurrence of secondary bulk separation. The criteria need to be broadened, as it is possible for secondary bulk separation to occur even if the eigenvector components are not equal. This is because the different surface tensions give rise to varying growth rates for each of the fluid component, and this has a complex interplay with the initial fluid concentrations. Similar to the equal surface tension criteria, we focus on the largest positive eigenvalue and its corresponding eigenvector. Following Eq. \eqref{eigenvector}, with $e_1$ and $e_2$ the components of the eigenvector with the largest positive eigenvalue, we consider two possible conditions for secondary bulk phase separation to be observed. First, if $C_1 = C_2$ and $e_1 = e_2$, then the two components grow together. This is the same condition as discussed in the previous section, but in fact this condition is rare to achieve when the surface tensions of the fluid interfaces are not equal. As before, without any loss of generality, we explicitly consider concentration variations in $C_1$ and $C_2$, and the third component can be obtained via $C_3 = 1 - C_1 - C_2$. The second possibility is if one component has a lower initial concentration ($C_1<C_2$), but its fluctuation has a faster growth rate $e_1 > e_2$ (or vice versa for $C_2<C_1$). In such a case, the two components are mixed together for some period before eventually phase separating. 
	
	Using this updated condition for the secondary bulk phase separation, along with the other conditions as described for equal surface tensions, we can compare the resulting phase diagrams. Fig. \ref{PhaseDiagrams} (a) shows the eigenvalue analysis, while (b) and (c) which represent the numerical and theoretical results respectively. As before, the eigenvalue analysis on its own has little predictive value for the morphologies observed in the simulations. Comparing Figs. \ref{PhaseDiagrams}(b) and (c), we further find that our phenomenological model extends to general surface tension values. It is clear that the key trends are captured for each phase separation mechanism, even though the boundaries of the different regions are less accurate when compared to the equal surface tension case.
	
	The deviations observed are mainly due to two reasons. First, at the boundary between binary-like and enrichment mechanisms, the minor component does not always enrich at the interface, especially when its total amount is very small compared to the simulation size. Here, the minor component remains mixed in the background components. Such tendency is more common with increasing surface tension, as it becomes more costly energetically to create interfaces. 
	
	Second, at the boundary between enrichment and secondary bulk phase separation mechanisms, there are several sources of uncertainties to classify the phase separation mechanism. In particular, the separation between the two more minor components may take place within a small droplet and before the components clearly reaching their expected bulk values. In addition, the condition $C_1 < C_2$ and $e_1>e_2$ must be considered with care. If $C_1$ is significantly smaller than $C_2$, then $e_1$ must be significantly larger than $e_2$ instead of being only modestly larger for secondary bulk separation to occur.  
	
\begin{figure}
		\centering
		\includegraphics[width=\linewidth]{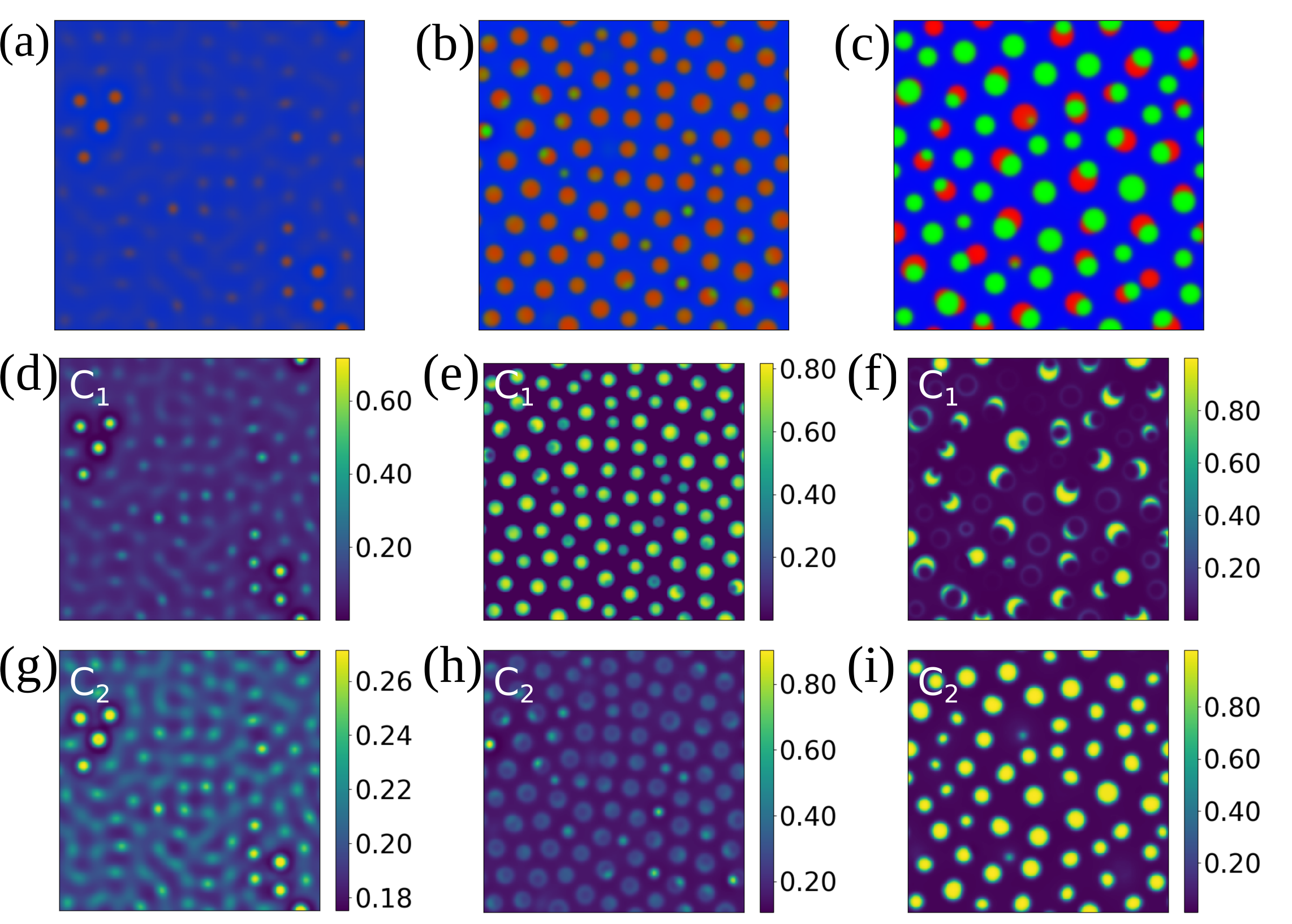}
		\caption{\label{SmallSeparation} Phase separation dynamics for initial concentrations: $C_1 = 0.1$ (red), $C_2=0.2$ (green), and $C_3=0.7$ (blue), with time evolving from left to right. Panels (a-c) show all components at once, while panels (d-f) and (g-i) only show $C_1$ and $C_2$.}
	\end{figure}

	To further highlight the importance of the eigenvectors in the phase separation dynamics, consider the results shown in Fig. \ref{SmallSeparation}. Here, we initialise the simulation with $A_1 = 0.1$ (red), $A_2=0.2$ (green), and $A_3=0.7$ (blue), and the normalised eigenvector corresponding to the largest positive eigenvalue is calculated to be $(e_1,e_2,e_3) = (-0.580,-0.208, 0.788)$. In agreement with our updated condition, both $C_1$ and $C_2$ remain mixed together for a significant length of time as the composite droplets emerge from the $C_3$ background. Then, despite the $C_1$ component being the most minor component, we observe it separates out before the second minor component ($C_2$). This is precisely because the eigenvector component for $C_1$ is twice larger compared to $C_2$. The component $C_2$ eventually begins to enrich at the interfaces to form patches.  
	
	Finally, we note that the surface tension values affect the possible morphologies, consistent with the observations in a recent work by Mao et al. \cite{mao2020designing}. For our choice of surface tensions in this section, the worm morphology is not possible during phase separation, unlike in the equal surface tension case. Since the Neumann angle $\theta_1 < 90^{\circ}$, the domains enriched in $C_1$ will form concave capillary bridges with negative pressure compared to their surroundings. These will lead to the surrounding domains quickly merging together, and as a result, the worm morphology cannot be supported.
	
	\section{Conclusion}
	
	In this work, we have systematically simulated spontaneous phase separation of ternary fluid mixtures across the composition space, both when all the fluid surface tensions are equal and when they all have different values. The observed possible phase separation mechanisms are in agreement with previous results \cite{nauman1994morphology, chen1994computer, huang1995phase}, and we have classified them into four basic types, which we term as (I) direct ternary phase separation, (II) primary and secondary bulk phase separation, (III) enrichment and instability at interface, and (IV) binary-like. Typically, these mechanisms lead to lattice, worm, patchy droplet and droplet morphologies.
	
	In contrast to binary phase separation, to understand the phase separation pathways, we have shown that the eigenvalues alone are insufficient, and it is important to consider both the eigenvalues and eigenvectors. Such consideration is in good agreement with direct simulation results, and it leads us to the conclusion that the enrichment and instability at interface mechanism is the dominant mechanism in ternary fluid phase separation. 
	
	We have also carried out preliminary work where we switch off the coupling to the Navier-Stokes equation, and where we vary the fluid viscosity ratios (up to a maximum of 10, data not shown) for the equal surface tension cases. We find the observed instability mechanisms are qualitatively very similar to the 4 reported here, and hence hydrodynamics effects will only lead to minor variations in the phase diagram. However, variations in the structures are found at the later stages of the coarsening. For example, the domains are more elongated in the diffusive regime without any coupling to hydrodynamics, akin to that reported for binary fluids \cite{wagner1998}. Indeed, in the future, it will be interesting to systematically study the possible scaling laws for the coarsening dynamics. For ternary fluid phase separation, we anticipate a much more complex growth laws. For example, in the primary and secondary bulk phase separation and the enrichment and instability at interface regimes, there are multiple length scales depending on whether we are interested in the majority or the minority fluid components.
	Furthermore, it will also be interesting to extend the study to three dimensional phase separation and to consider more than three fluid components, including when the components are all immiscible and when they are partially miscible. When there are four or more fluid components, an open question is whether the enrichment in the enrichment and instability mechanism remains occurring at the interface, or if it favours the junction of three or more domains.
	
	\section*{Conflicts of interest}
	There are no conflicts to declare
	
	\section*{Acknowledgements}
	ACMS is supported by the Engineering and Physical Sciences Research Council's Centre for Doctoral Training in Soft Matter and Functional Interfaces (grant no. EP/L015536/1). HK acknowledges funding from the Engineering and Physical Sciences Research Council (grant no. EP/V034154/1).
	
	All data supporting this study are available from the authors upon reasonable request.
	
	\bibliographystyle{rsc}
	\bibliography{Ref}

\end{document}